\documentclass[final,5pt,twocolumn]{elsarticle}
\usepackage{lineno,hyperref}
\modulolinenumbers[5]
\usepackage[utf8]{inputenc}
\usepackage{amsmath}
\usepackage{amsfonts}
\usepackage{amssymb}
\usepackage{lmodern}
\usepackage{xfrac}
\usepackage{graphicx}
\usepackage{subfig}
\graphicspath{ {./figures/} }
\usepackage{fullpage}
\usepackage{float}
\usepackage{caption}
\usepackage{siunitx}
\usepackage{xcolor}

\biboptions{sort&compress}

\begin{document}

\begin{frontmatter}

\title{Formation of Nb-rich droplets in laser deposited Ni-matrix microstructures}

\author[mymainaddress]{Supriyo Ghosh\corref{mycorrespondingauthor}}
\ead{supriyo.ghosh@nist.gov}
\author[mymainaddress]{Mark~R.~Stoudt}
\author[mymainaddress]{Lyle~E.~Levine}
\author[mymainaddress]{Jonathan~E.~Guyer\corref{mycorrespondingauthor}}
\cortext[mycorrespondingauthor]{Corresponding author.}
\ead{jonathan.guyer@nist.gov}
\address[mymainaddress]{Materials Science and Engineering Division, National Institute of Standards and Technology, Gaithersburg, MD 20899, USA}

%


\begin{abstract}
Ni-rich $\gamma$ cells and Nb-rich eutectic droplets that form during laser power bed fusion solidification of Ni-Nb alloys are studied using experiments and simulations. Finite element simulations estimate the cooling rates in the melt pool and phase-field simulations predict the resulting cellular microstructures. The cell and droplet spacings are determined as a function of cooling rate and fit to a power law. The formation of Laves phase is predicted for a critical composition of Nb in the liquid droplets. Finally, our simulations demonstrate that anisotropy in the $\gamma$ orientation influences the Laves fraction significantly. 
\end{abstract}


\end{frontmatter}
Ni-based superalloys possess excellent mechanical properties and corrosion resistance up to high temperatures primarily due to the fine precipitation of Nb-rich phases and are therefore used in gas-turbine and jet-engine components \cite{Attallah2016}. The laser powder bed fusion (LPBF) additive manufacturing (AM) process is used to fabricate or repair these alloys by layer-by-layer application of the alloy powder and subsequent repeated melting, solidification and solid-state phase transformations~\cite{Murr2012,King2015,Francois2017,Trevor2017}. The solidification in this process often results in a columnar face-centered-cubic $\gamma$-Ni matrix and microsegregation of Nb, Mo and Ti in the interdendritic regions~\cite{Murr2012,Nandwana2017,Tian2017,Zhang2017_nist}. The regions with high concentration of Nb often transform to intermetallic phases during terminal solidification. One of those phases is the Laves phase. Laves drastically reduces the tensile strength, fracture toughness and low-cycle fatigue properties of the additively manufactured material. An understanding of the formation and control of Laves is therefore essential.

Under nonequilibrium solidification conditions, solute redistribution across a solid-liquid interface during the growth of primary $\gamma$ phase leads to severe Nb segregation in the liquid molten pool~\cite{kurzbook}. During terminal solidification, as the roots of the $\gamma$ cells coalesce in the semisolid mushy zone, Nb-rich liquid channels between the cells are separated into isolated droplets~\cite{supriyo20173d}. It is difficult to entirely avoid the formation of these droplets due to the rapid nature of cooling during LPBF solidification. The metastable liquid in the form of droplets could potentially undergo a nonequilibrium reaction below the eutectic temperature and transform to a combination of Laves and $\gamma$. Since Laves is brittle and makes the as-deposited microstructures weak, there have been several experiments \cite{Xiao2017,Zhang2013,Antonsson2005,Ram2004,Guo2017} and simulations \cite{nie2014} to suggest approaches to minimize its formation. The most widely used approach is homogenization heat treatment \cite{Radhakrishna1997,Zhao2008,Qi2009,Ram2004,Zhang2017_nist}. A manipulation of the solidification conditions in the melt pool by heat input/cooling rate was also found to be effective in controlling the morphology and distribution of Laves. High cooling rates resulted in a fine and discrete Laves network beneficial for mechanical properties, whereas low cooling rates resulted in a coarse and continuous network detrimental for the same \cite{Xiao2017,nie2014,Qi2009,Ram2005}. Laves was found to be refined significantly and reduced/eliminated in ultrarapid cooling rates \cite{Antonsson2005,Qi2009,Zhang2013}. The morphological transition of $\gamma$ phase from columnar to equiaxed, due to an increase in the cooling rate, was found effective for separating a continuous liquid network into isolated droplets/Laves \cite{nie2014,Ram2005}. While previous studies considered cooling rates on the order of $10^3$~K s$^{-1}$, the present work uses cooling rates on the order of $10^6$~K s$^{-1}$, consistent with LPBF. The microstructure-property correlation is therefore expected to be different than that reported in the existing literature. In what follows, we present the as-deposited microstructures from LPBF experiments and finite element and phase-field simulations.

\begin{figure}[h]
\begin{center}
\includegraphics[scale=0.18,angle=270]{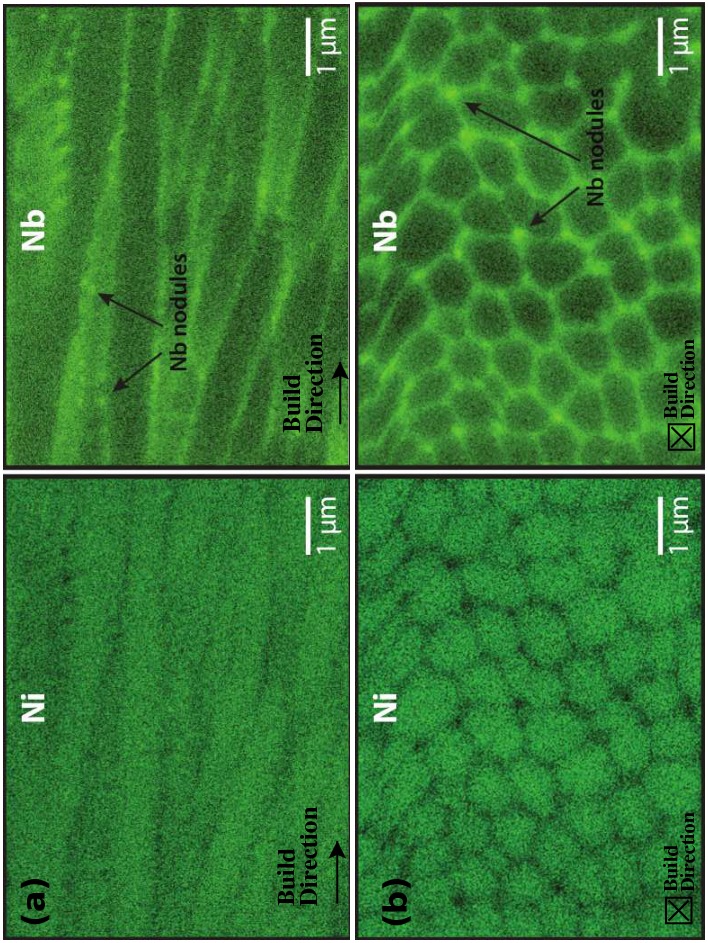}
\caption{(a) SEM/EDS maps in the build direction reveal the formation of Ni-rich $\gamma$ columnar microstructure. Nb-rich segregation is observed in the interdendritic regions. (b) The columnar $\gamma$ cells/dendrites and the interdendritic Nb segreation are presented in the direction perpendicular to the build direction. Note that the sparsely distributed bright spots are considered as Nb-rich droplets which could potentially evolve in time and transform to ($\gamma$ $+$ Laves) eutectic.}\label{fig_experiment}
\vspace{-8mm}
\end{center}
\end{figure}

Fifteen millimeter cubes were additively produced from virgin Inconel 625 alloy powder using an EOS M270 LPBF system\footnote{Any mention of commercial companies or products herein is for information only; it does not imply recommendation or endorsement by NIST.}. The standard EOS parameter set for this alloy was used consisting of a laser power of 195 W, scan speed of 800 mm s$^{-1}$, nominal powder layer thickness of \SI{20}{\micro \meter} and hatch spacing of \SI{100}{\micro \meter}. The samples were then cut from the build plate with electro-discharge machining in the as-built condition, i.e. the specimens did not undergo any stress-relief heat treatment. Portions of the as-built material were mounted and polished using standard metallographic techniques for scanning electron microscopy (SEM) analysis \cite{Vander1999}. The samples were etched via immersion in aqua regia for 10 s to 60 s to reveal the microstructure. A final polishing step using a vibrational polishing system with \SI{0.2}{\micro \meter} colloidal silica was employed to provide a strain-free surface for SEM electron backscatter diffraction and energy dispersive X-ray spectroscopy (EDS). The as-built microstructures from the EDS spectra consist of Ni-rich $\gamma$ matrix and Nb, Mo, C and other elemental segregation \cite{Eric2017, Zhang2017_nist}. We consider a binary analog of these microstructures, i.e. Ni matrix and Nb segregation, to describe the microstructural evolution. The as-built microstructures are presented in Fig.~\ref{fig_experiment}. These consist of primary Ni-rich $\gamma$ cells/dendrites (average spacing $\approx \SI{0.6}{\micro \meter}$) and Nb-rich interdendritic regions. Although not clear, the secondary/tertiary sidearms cut the interdendritic space into smaller regions, and left less space for Nb-rich spots to grow in a sphere-like morphology. These spots appear bright and are extremely fine; the average spacing is $\approx \SI{0.26}{\micro \meter}$, average diameter is $\approx \SI{0.13}{\micro \meter}$, and average area fraction is $\approx$ \SI{2}{\%} to \SI{3}{\%}. The concentration of these spots could not be resolved since the beam spot size was quite large with respect to the size of the segregation features. 

\begin{figure*}[ht]
\begin{center}
\includegraphics[scale=0.25]{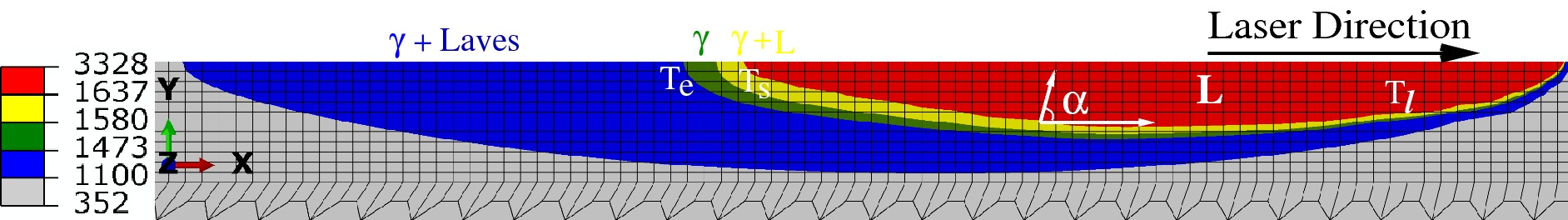}
\caption{The temperature distribution along a 2D section cut along the centerline of a simulated 3D melt pool is presented. To roughly correspond with the Ni-\SI{5}{\%} Nb phase diagram~\cite{knorovsky1989,Nastac1996}, red represents the liquid phase ($L$), yellow represents the mushy zone ($\gamma + L$), green represents the solid $\gamma$ phase, and blue represents the eutectic ($\gamma$ + Laves) existence. $T_l$ = 1637 K, $T_s$ = 1580 K and $T_e$ = 1473 K are the liquidus, solidus and eutectic temperatures, respectively. The solidification conditions are estimated from the red-yellow liquidus boundary. The temperature gradient $G$ is estimated by its magnitude $|\nabla T|$. The solidification velocity $V$ is estimated by $V_b \cos \alpha$, where $V_b$ is the beam speed and $\alpha$ the local solidification angle.}\label{fig_meltpool}
\vspace{-8mm}
\end{center}
\end{figure*}

The solidification conditions in the above experiment were estimated by heat transfer finite element simulations and reported in previous works by us and our collaborators~\cite{ Brandon2015,Trevor2017,supriyo2017}. Here we show the typical temperature distribution during this LPBF simulation in Fig.~\ref{fig_meltpool}. Referring to this temperature profile, the $\gamma$ cells/dendrites solidify directionally and grow perpendicular to the solid-liquid boundary approximated by $T_l$ isotherm in a temperature gradient $G$ and at a solidification velocity $V$. The solid-liquid growth front represents different $G$ and $V$. We note that $G$ ranges from $ 2.4 \times 10^7$ K m$^{-1}$ to $0.14 \times 10^7$ K m$^{-1}$ and $V$ ranges from 0.01 m s$^{-1}$ to 0.3 m s$^{-1}$ as we move from the bottom to the rear of this boundary. $G$ is translated along the build direction ($z$) by the pulling velocity $V$ in a directional ``frozen temperature'' solidification framework for microstructure evolution. $G$ times $V$ is the cooling rate $\dot{T}$.


We use a phase-field model detailed in Refs.~\cite{Trevor2017,supriyo20173d} where the first simulations of $\gamma$ cells during solidification of a dilute Ni-Nb alloy, a binary approximation of a Ni-based superalloy, were reported. A conserved composition field $c$ and a non-conserved phase-field variable $\phi$ are used to label the microstructure phases; $\phi = 1$ in the solid, $\phi = - 1$ in the liquid, and the solid-liquid interface is automatically extracted by the contour $\phi = 0$. An antitrapping solute flux~\cite{Karma2001, Echebarria2004} was introduced to minimize the interface-induced solute partitioning at low $\dot{T}$ leading to effective solute rejection in the liquid in front of the advancing cells. The effects of melt convection are not included in this model and solute is transported in the liquid by diffusion only. The time-dependent $\phi$ and $c$ equations of motion are solved on a uniform mesh, using the finite volume method, explicit time stepping scheme and zero-flux boundary conditions. The size of the simulation box in the growth ($z$) direction is \SI{40}{\micro\metre}, and a representative domain size, $L_x \times L_y$, \SI{4}{\micro\metre}$\times$\SI{4}{\micro\metre} is used for 3D simulations. Other numerical and thermophysical parameters are detailed in Refs.~\cite{supriyo2017,supriyo20173d}. In this parameter space, our simulation results presented below become virtually independent of the discretization size (\SI{8}{nm}) and the interface thickness (\SI{10}{nm}) values.

\begin{figure}[ht]
\begin{center}
\includegraphics[scale=0.2,angle=270]{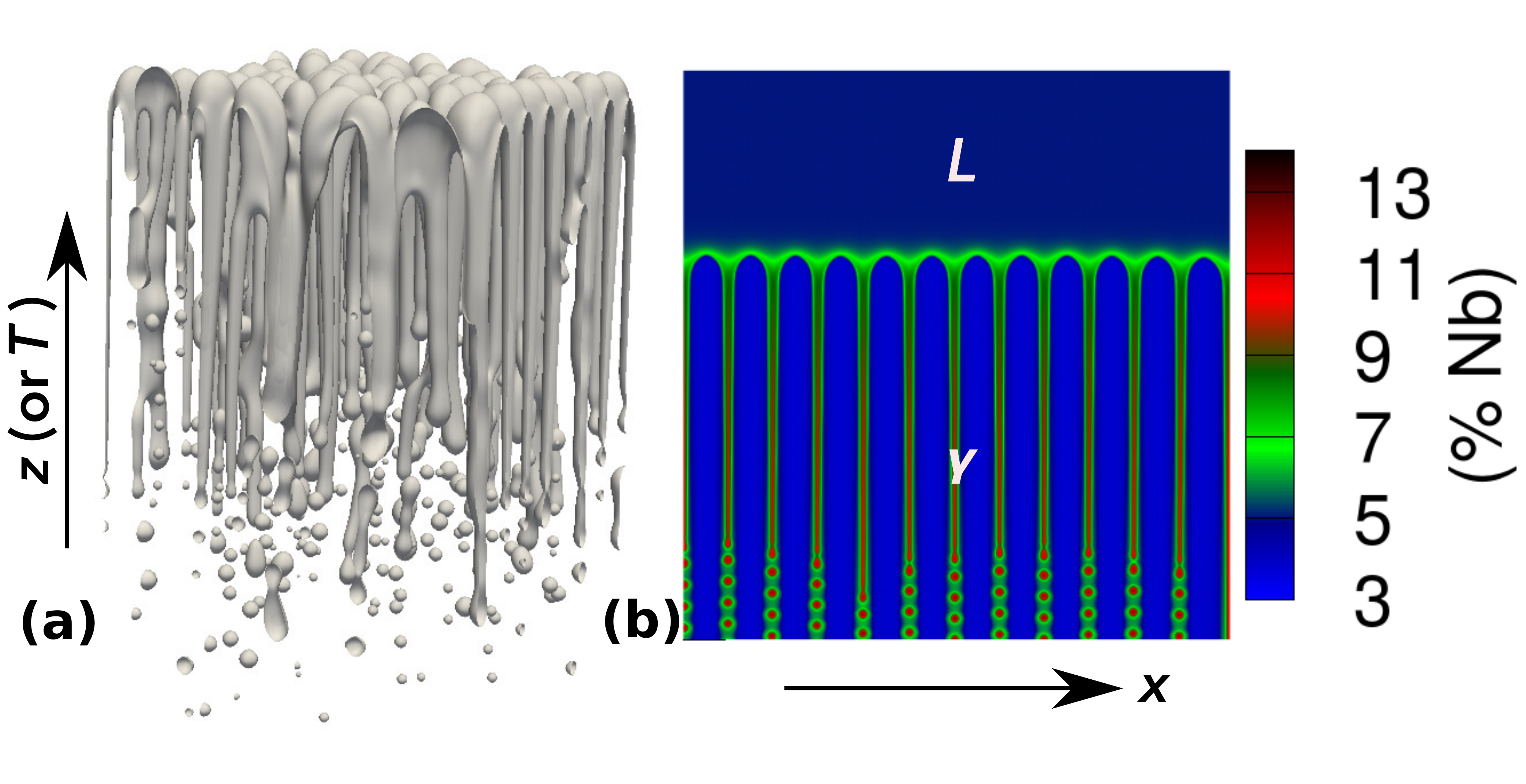}\vspace{-2mm}
\caption{Phase-field simulations begin with a thin layer of solid at the bottom of the simulation box with initial Nb compositions in the solid and the liquid, as in~\cite{supriyo2017}. Small, random amplitude perturbations are applied at the initial solid-liquid interface, from which stable perturbations grow with time and break into steady state $\gamma$ cells. (a) Snapshot picture of 3D steady state cellular growth front for $\dot{T}$ = $5 \times 10^5$ K s$^{-1}$ extracted at the contour $\phi = 0$. Temperature gradient or build direction is vertical. Nb-rich droplets pinch off of the cell roots. The average cell spacing $\lambda_c$ = \SI{0.22}{\micro\metre} and droplet spacing $\lambda_d$ = \SI{0.13}{\micro\metre}. (b) Spatial distribution of Nb across 2D cells is represented with clear visualization of the diffusion length (green); here, $\lambda_c$ = \SI{0.38}{\micro\metre} and $\lambda_d$ = \SI{0.22}{\micro\metre}.}\label{fig_figure}
\vspace{-5mm}
\end{center}
\end{figure}

The solidification pathway for a Ni$-$\SI{5}{\%} Nb\footnote{Concentration is represented in mass fraction.} alloy is given by $L$ $\rightarrow$ $L$ + $\gamma$ $\rightarrow$ $\gamma$ + Laves ~\cite{knorovsky1989,Nastac1996}. These phase transformations occur due to different degrees of undercooling below $T_l$, and the resultant microstructures are predicted by phase-field simulations in Fig.~\ref{fig_figure}. The microstructures consist of Ni-rich $\gamma$ cells and Nb-rich intercellular regions and correspond only to a particular position along the melt pool boundary. The average distance between the $\gamma$ cells remains constant in steady state, which is the cell spacing or the primary dendrite arm spacing. As the cells grow in the liquid, Nb is rejected through the cell-liquid interface in a nonequilibrium partitioning process~\cite{supriyo2017,Trevor2017}. Nb thus varies in the liquid ahead of the cell tip, in the liquid between cells, and in the solid cell core, resulting in complex segregation features (Fig.~\ref{fig_figure}b). The microsegregation or the composition gradient between the cell core and the periphery of individual cells is extracted by a composition-distance profile across the cells and reported in \cite{supriyo2017}. The rejection of Nb by the growing cells increases the Nb content in the liquid. During terminal solidification, close to the bottom of the simulation box, as the roots of the solid cells grow toward each other and coalesce in the mushy zone at a low temperature, Nb-rich liquid in the intercellular channels is separated into isolated droplets, as in~\cite{Ma2014, supriyo2017, ungar1985, Boettinger1999}. Since the diffusion path is absent at lower temperatures, the Nb content in these droplets increases rapidly with a reduced residual liquid fraction with increasing distance below the cellular growth front. These droplets could undergo eutectic transformation beyond a threshold composition of Nb, resulting in ($\gamma$ + Laves) eutectic. The present binary model does not represent any phase beyond $L$ and $\gamma$. Therefore, the formation of Laves is predicted using a criterion for the threshold composition of Nb in the liquid. Different critical values of Nb were reported in literature to describe the Laves formation. Dupont~\emph{et al.} considered the Laves formation due to a Nb composition in the liquid $>$ \SI{23}{\%}, whereas Nastac and Stefanescu \cite{Nastac1996} and Peng~\emph{et al.} \cite{Peng2013} used a value of \SI{19}{\%} to describe the same. We consider the criterion used by Nie~\emph{et al.} \cite{nie2014} and Antonsson~\emph{et al.} \cite{Antonsson2005}. In this approach, the liquid with Nb $>$ \SI{20}{\%} transforms into Laves phase and the liquid with Nb $\leqslant$ \SI{20}{\%} transforms into $\gamma$ phase.

\begin{figure*}[ht]
\begin{center}
\includegraphics[scale=0.27,angle=270]{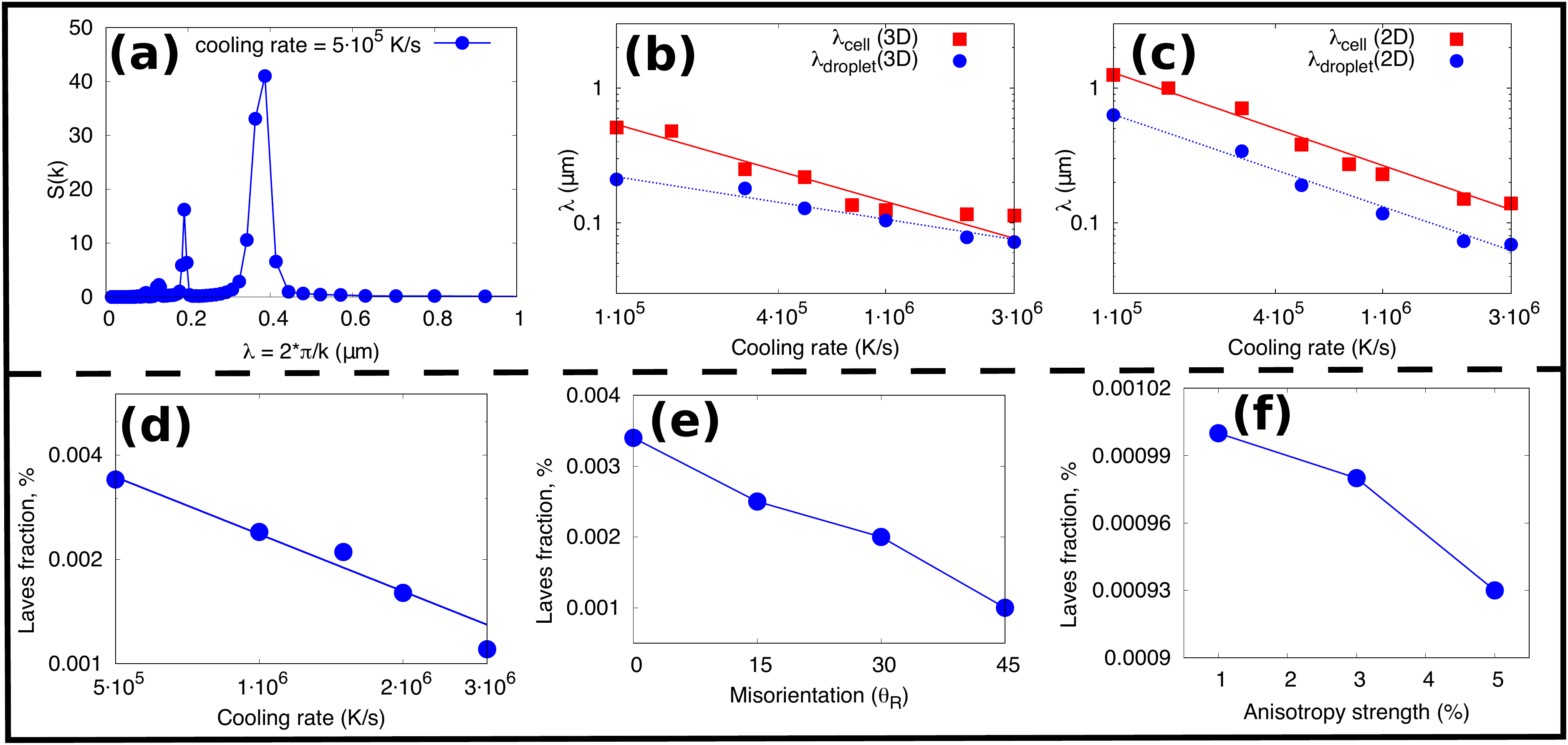}\vspace{-2mm}
\caption{(a) The power spectrum is presented for the 2D cellular structure in Fig.~\ref{fig_figure}b. The main peak corresponds to the average cell spacing $\lambda_c$. Second highest peak is the average droplet spacing $\lambda_d$. (b) Simulated $\lambda_c$ and $\lambda_d$ values in 3D are plotted against cooling rates with a fit to $\lambda_c$ and $\lambda_d$ =  $A(\dot{T})^n$ in $\log$ scale. (c) $\lambda_c$ and $\lambda_d$ values from 2D simulations are presented with the same approach for the 3D data. (d) The fraction of Laves, $f_d$, decreases with increasing logarithmic cooling rates. (e) The solid-liquid (cubic) anisotropy function in a plane in the growth direction can be modeled by $a(\theta) = 1 + \epsilon_{sl} \cos (\theta-\theta_R)$, where $\epsilon_{sl}$ is the anisotropy parameter, $\theta$ is the angle between the interface normal and horizontal direction in the laboratory reference frame and $\theta_R$ is an in-plane rotation angle. $f_d$ decreases with increasing misorientation $\theta_R$ between the growth direction and build direction. (f) $f_d$ decreases with increasing $\epsilon_{sl}$ (or decreasing the minimum of the solid-liquid interface energy).
}\label{fig_spacing}
\vspace{-8mm}
\end{center}
\end{figure*}

The spacing between the $\gamma$ cells as well as between the droplets is related to the yield and tensile strengths of the solidified material. Prediction and control of the spacing between $\gamma$ and droplets are therefore essential. We extract the average cell and droplet spacings from the simulated cellular microstructures by the calculation of the mean power spectrum: $S(k) = |h(k)|^2$, where $h(k)$ is the Fourier transform of the solid-liquid interface profile $h(z)$ and $k$ is the wave number. From this analysis, $\lambda_c$ and $\lambda_d$ are estimated by $\lambda_c$ (or $\lambda_d$) = $2\pi/k_{mean} = \sum_{k>0} kS(k) / \sum _{k>0} S(k)$, as shown in Fig.~\ref{fig_spacing}a. The highest peak in this spectrum corresponds to the dominant wavelength in the microstructure, i.e. $\lambda_c$. For the simulated cooling rates, the estimated $\lambda_c$ ranges from \SI{0.1}{\micro\metre} to \SI{0.5}{\micro\metre} in 3D (Fig.~\ref{fig_spacing}b) and from \SI{0.2}{\micro\metre} to \SI{1.1}{\micro\metre} in 2D (Fig.~\ref{fig_spacing}c). Solute rejection/diffusion at the cell tip is more efficient in 3D than that in 2D; $\lambda_c$ is therefore smaller in 3D than in 2D. The simulated cell spacing data agree reasonably with our experiment, where $\lambda_c$ is estimated between \SI{0.5}{\micro\metre} and \SI{1.0}{\micro\metre}. Similar observations were also made by Amato \emph{et al.}~\cite{Amato2012} for Inconel alloys.
The second dominant wavelength in Fig.~\ref{fig_spacing}a is the droplet spacing $\lambda_d$. In our simulations, $\lambda_d$ ranges from \SI{0.1}{\micro\metre} to \SI{0.2}{\micro\metre} in 3D (Fig.~\ref{fig_spacing}b) and from \SI{0.1}{\micro\metre} to \SI{0.6}{\micro\metre} in 2D (Fig.~\ref{fig_spacing}c). The droplets are finer in 3D than that in 2D due to the same above reason for $\lambda_c$. Note that the average $\lambda_d$ estimated from the experiment is $\approx \SI{0.26}{\micro \meter}$, which compares reasonably with our simulation data. Since the droplets form after the cell roots have coalesced, resulting in a large $\gamma$ fraction in the mushy zone, and we ignore diffusion in the solid, their relative position does not evolve with time any more. The average diameter of the droplets at this stage is estimated between \SI{25}{\nano\meter} and \SI{200}{\nano\meter} with an error on the order of \SI{10}{\%}, in good agreement with our experiment ($\approx \SI{130}{\nano \meter}$).

In the literature, $\lambda_c$ is often described by a power law: $\lambda_c =  A(\dot{T})^n$, where $A$ and $n$ are material constants. Since both $\lambda_c$ and $\lambda_d$ are established in the microstructure due to the same segregation event, the same power law is used to represent both. The line of best fit representing the power law is drawn through our simulation data in Figs.~\ref{fig_spacing}b and \ref{fig_spacing}c. It is evident that both $\lambda_c$ and $\lambda_d$ decrease with increasing $\dot{T}$. We catalog the fitting values of $A$ and $n$ for a reference which can be compared to other studies. The $\lambda_c$ data are fitted best with $A = \SI{800}{\micro\metre}$ and $n = - 0.6$ in 2D and $A = \SI{392}{\micro\metre}$ and $n = - 0.6$ in 3D. Note that the spacing selection (at the steady state growth front resulting in $\lambda_c$) follows a similar slope ($n$) in both 2D and 3D, however, $\lambda_c$ in 3D is about twice that in 2D. On the other hand, the $\lambda_d$ data are drastically different between 2D and 3D; $A = \SI{1672}{\micro\metre}$ and $n = - 0.7$ in 2D while $A = \SI{8}{\micro\metre}$ and $n = - 0.3$ in 3D. One likely reason for these differences could be that the droplet formation in experiments and 3D simulations is noisy, which is not surprising, due to apparent randomness during rapid solute redistribution due to rapid cooling rates, leading to random/rapid joining of the cell roots in the semisolid mushy zone and random/rapid droplet pinch off events close to cell roots. The microstructural features in the mushy zone thus never reach a true steady state in either experiments or in 3D simulations~\cite{Space2001, Ma2014}.

Since the solidification conditions control the Nb partitioning and hence segregation in the melt pool, droplet formation is strongly dependent on these conditions. Previous experiments reported that the volume fraction of droplets decreases with increasing cooling rate, and droplet formation was inhibited in ultrarapid cooling rates~\cite{Qi2009,Zhao2008,Zhang2013}. A higher $\dot{T}$ increases the solid-liquid interface growth speed and decreases the degree of microsegregation, giving insufficient time for Nb to diffuse from the cell core to the liquid. As a result, more Nb is trapped within $\gamma$ cells and less Nb is available to form droplets. We note that the average volume fraction of Nb-rich droplets from 3D simulations is between \SI{2.1}{\%} and \SI{3.4}{\%}, in good agreement with our experiment. These droplets eventually phase transform to ($\gamma$ $+$ Laves) eutectic in solid-state. We estimate the average volume fraction of Laves ($f_d$) on the basis of Nb $>$ \SI{20}{\%} limit and present the influence of $\dot{T}$ on $f_d$ in Fig.~\ref{fig_spacing}d. On average, $f_d$ decreases with increasing $\dot{T}$. For the $\dot{T}$ values used in our simulations, $f_d$ varies between \SI{0.001}{\%} and \SI{0.004}{\%}. Due to such a small fraction of $f_d$, diffraction peaks of Laves phase could not be detected by XRD analysis~\cite{Liu2013, Eric2017}. We wish to note that certain differences may exist between simulation and experiment since we use a strongly idealized finite element model (ignoring melt convection, Marangoni flow and other hydrodynamic effects~\cite{King2015} in the melt pool) to estimate $\dot{T}$. Nie~\emph{et al.} \cite{nie2014} considered $\dot{T}$ on the order of 10$^3$ K s$^{-1}$ and estimated $f_d$ between \SI{5}{\%} and \SI{10}{\%}, using 2D stochastic analysis simulations. The experiments of Cieslak~\emph{et al.} \cite{Cieslak1990} and Ram~\emph{et al.} \cite{Ram2005} also predicted $f_d$ between the above limits at a lower $\dot{T}$, whereas Ling~\emph{et al.} \cite{Ling2015} predicted a maximum $f_d$ of \SI{2}{\%} even at a lower $\dot{T}$. Our phase-field simulations are conducted at a higher $\dot{T}$ and resulted in a lower $f_d$. The variation of $f_d$ with $\dot{T}$ is found to fit best in a logarithmic scale with the form $f_d = a(\dot{T})^b$, where $a$ and $b$ are material constants. The line of best fit through $f_d$ vs. $\dot{T}$ data yields $a = \SI{4.6}{\%}$ and $b = - 0.6$.


It is generally accepted that $\gamma$ cells grow in the direction of thermal gradient. The orientation/texture of these cells however can be different in different locations in a solidifying molten pool~\cite{Liu2013,Nandwana2017,Attallah2016}. A slight deviation, for instance, of the scanning path of the laser beam can markedly change the spatial and temporal solute redistribution across the cell-liquid interface. Anisotropy in orientation, that is, change in the preferred growth direction with respect to the build direction, leading to intrinsic anisotropy in the mechanical properties, is therefore natural to consider. The distribution/fraction of droplets could therefore be a function of the orientation of the $\gamma$ cells. This issue has received little attention, but it may be important to consider during LPBF solidification. One simple way to test this using phase-field simulation is to vary the solid-liquid boundary anisotropy and the interface rotation during the growth of $\gamma$ cells. As a result, cells grow at an angle with the build (vertical) direction, and the $f_d$ becomes a decreasing function of the misorientation between the growth direction and the build direction, as shown in Fig.~\ref{fig_spacing}e. Note that the $f_d$ is calculated for zero misorientation in Fig.~\ref{fig_spacing}d, which varies significantly for angular growth directions. To further illustrate the different $f_d$ values for different growth conditions, the surface energy anisotropy in the solid-liquid boundary is varied between \SI{1}{\%} and \SI{5}{\%} in our simulations. Note that we have not noticed any qualitative changes in the cellular features such as the tip radius with respect to the changes in $\gamma$-anisotropy. On average, $f_d$ decreases with increasing magnitudes of anisotropy (Fig.~\ref{fig_spacing}f), which is desired. The misorientation and solid-liquid interfacial anisotropic properties may therefore be engineered to control the fraction of droplets and hence Laves in the microstructure. 
 

We ignore the effects of melt convection on the cell and droplet spacings and the formation of droplets. Effects of convection on the primary dendrite spacing are not as pronounced as compared to the secondary arms~\cite{Wang2003, Lee2010}, which are not observed in our simulations. Interestingly, the phase-field simulations by Lee \emph{et al.}~\cite{Lee2010} showed that the effects of convection are negligible in 3D simulations and thus the solute redistribution across the cell-liquid interface remained similar when simulations were conducted with and without convection. The simulated cells and droplets are extremely fine and provide significant resistance to fluid flow following an exponential increase of the damping effect in the mushy region, leading to reduced effects of convection~\cite{Yang2001,Tan2011}. In addition, consideration of a dilute alloy (\SI{5}{\%} Nb) reduces the effects of convection on Nb~\cite{Wang2003}. Therefore, phase-field simulations have been performed with reasonable approximations to predict the average cell and droplet spacings and droplet formation. A multicomponent phase-field framework \cite{supriyo2014,Ghosh2015,Ghosh2017_eutectic} will be used in future to represent $\gamma$, liquid and Laves phases in the microstructure, and melt convection will be considered for more accurate microstructure evolution.



\section*{Acknowledgements}
S. G. thanks Li Ma for finite element simulations, Maureen E. Williams for help in experiments, Gregor Jacob and Christopher Brown for preparing the AM builds, and Kevin McReynolds, Nana Ofori-Opoku and Sukriti Manna for useful discussions.
\section*{References}
\bibliography{papers}

\end{document}